\documentclass[]{JHEP3}
\usepackage{epsf,epsfig,subfigure,axodraw,graphicx,amsmath,amssymb}

\title{\Large\bf PAMELA/ATIC anomaly \\
from the meta-stable extra dark matter component \\ and the
leptophilic Yukawa interaction }

\author{\large
Bumseok Kyae\\
Department of Physics and Astronomy and Center for
Theoretical Physics, Seoul National University, Seoul 151-747\\
E-mail: \email{bkyae@phya.snu.ac.kr} }

\abstract{

We present a supersymmetric model with two dark matter (DM)
components explaining the galactic positron excess observed by
PAMELA/HEAT and ATIC/PPB-BETS: One is the conventional (bino-like)
lightest supersymmetric particle (LSP) $\chi$, and the other is a
TeV scale meta-stable neutral singlet $N_D$, which is a Dirac
fermion ($N,N^c$). In this model, $N_D$ decays dominantly into
$\chi e^+e^-$ through an $R$ parity preserving dimension 6
operator with the life time $\tau_N\sim 10^{26}$ sec. We introduce
a pair of vector-like superheavy SU(2) lepton doublets ($L,L^c$)
and lepton singlets ($E,E^c$). The dimension 6 operator leading to
the $N_D$ decay is generated from the leptophilic Yukawa
interactions by $W\supset Ne^cE+Lh_dE^c+m_{3/2}l_1L^c$ with the
dimensionless couplings of order unity, and the gauge interaction
by ${\cal L}\supset \sqrt{2} g'\tilde{e}^{c*}e^c\chi~+~$h.c. The
superheavy masses of the vector-like leptons ($M_L, M_E\sim
10^{16}$ GeV) are responsible for the longevity of $N_D$. The low
energy field spectrum in this model is just the MSSM fields and
$N_D$.
Even for the case that the portion of $N_D$ is much smaller than
that of $\chi$ in the total DM density [${\cal O}(10^{-10})
\lesssim n_{N_D}/n_\chi$], the observed positron excess can be
explained by adopting relatively lighter masses of the vector-like
leptons ($10^{13}$ GeV $\lesssim M_{L,E} \lesssim 10^{16}$ GeV).
The smallness of the electron mass is also explained. This model
is easily embedded in the flipped SU(5) grand unification, which
is a leptophilic unified theory.

}

 \preprint{SNUTP 09-001}

\begin{document}

\section{Introduction}

Although the existence of dark matter (DM)
is advocated through several cosmological observations, we don't
know yet its identity.
%
%
The lightest supersymmetric particle (LSP) in the minimal
supersymmetric standard model (MSSM) has been believed to be one
of the most promising DM candidates over last two decades. Not
only it is well-motivated from the promising particle physics
model i.e. the MSSM, but also it naturally carries the features
required for a weakly interacting massive particle (WIMP).

However, recently the PAMELA group reported very challenging
observational results on excess of high energy positrons coming
from the galactic halo \cite{PAMELA}. It has confirmed the
previous similar observations of HEAT \cite{HEAT} with the
improved precisions. The PAMELA data shows a rising positron flux
$e^+/(e^++e^-)$ from 10 to 100 GeV, which gives rise to
considerably big deviations ($\sim 0.1$) from the theoretically
expected values~\cite{Moskalenko:1997gh}. On the other hand, any
significant anti-proton excess was not observed. Indeed, these
observations would be quite hard to explain in large classes of
models only with one DM component of the Majorana fermion such as
the MSSM.
Moreover, the ATIC and PPB-BETS collaborations also reported
recently their observations that the flux of $(e^++e^-)$ keeps
rising upto around $800$ GeV \cite{ATIC,PPB-BETS}. It might imply
that if such deviations result from the DM's annihilation or
decay, the mass of the DM would be in a TeV range.

Various scenarios beyond the conventional DM models have been
suggested so far, including new scenarios of DM annihilation
\cite{HKK,BHKKV,annihilation}, DM decay
\cite{decay,strumia,Hamaguchi}, and nearby pulsars producing
charged leptons \cite{pulsar}.
However, if the mass of DM is above TeV, and if we keep the
galactic profile of NFW and Einasto \cite{NFW}, the annihilation
scenario would be disfavored \cite{aticGammaray} due to the bound
of the $\gamma$ ray flux by the HESS observations of the galactic
ridge \cite{HESS}. (See also the recent discussions on DM
annihilation in Refs. \cite{aticAnnihil}.)

In Refs.~\cite{HKK,BHKKV}, the PAMELA/HEAT anomaly could be
explained by co-annihilations of the LSP and another DM component
$N_D$, which is a Dirac fermion $(N,N^c)$ with a weak scale mass.
It turns out that in this case, introductions of a pair of extra
vector-like lepton singlets $(E,E^c)$, and a leptophilic coupling
of the DM, $Ne^cE$ are indispensable to explain the PAMELA/HEAT's
observations. The relatively small masses of $(N,N^c)$ and
$(E,E^c)$ [$\sim {\cal O}(100)$ GeV] and the needed $N^3$ coupling
in the superpotential for decay of $E$ and $E^c$ into the DM can
be guaranteed by the U(1)$_R$ symmetry. This model (``$N_{\rm
DM}$MSSM'') is easily embedded in the flipped SU(5) grand unified
theory (GUT), which is assumed to be broken to the MSSM at the GUT
scale \cite{BHKKV,Barr82}.

In this paper, we attempt to explain the anomalies from
PAMELA/HEAT and ATIC/PPB-BETS by {\it decay} of the extra DM
component $N_D$. If the observations by PAMELA/HEAT and
ATIC/PPB-BETS are caused indeed by DM decay, the data of
ATIC/PPB-BETS imply that the mass of the DM matter would be around
2 TeV and its life time $\sim 10^{26}$ sec. Such a long life time
is achievable, if the TeV scale DM decays to $e^\pm$ (+ neutral
particles) through a dimension 6 operator suppressed by the mass
squared of order the GUT scale \cite{strumia,Hamaguchi}. To be
consistent with the PAMELA data, the hadronic decay modes of DM
should not exceed 10 $\%$ \cite{strumia}.

This paper is organized as follows: In Sec. 2, we discuss how a
dimension 6 operator for DM decay can be naturally dominant in a
supersymmetric model extending the MSSM, and in Sec. 3, we propose
a specific model realizing the idea discussed in Sec. 2. In Sec.
4, we briefly discuss the implications of the recently observed
anomalies in phenomenological model building in string theory, and
summarize our conclusions.

\section{Dark Matter Decay}

For the case that DM decays to the positrons, the positron flux at
the earth is given by a semi-analytic form \cite{strumia}:
\begin{eqnarray} \label{flux}
\Phi_{e^+}(E)=\left(\frac{\rho}{m_{\rm DM}}\right)\Gamma_{\rm
DM}\times \frac{1}{4b(E)}\int^{m_{\rm
DM}}_{E}dE'~\frac{dN_{e^+}}{dE'}~ I(\lambda_D) ,
\end{eqnarray}
where $dN_{e^+}/dE'$ is the spectrum of $e^+$ produced by DM
decay, $\rho$ the DM energy density at the earth, and $m_{\rm DM}$
its mass. $b(E)$ [$=E^2/$(GeV$\cdot 10^{16}$ sec.)] indicates the
energy loss coefficient, and $I(\lambda_D)$ is the halo function
for DM decay, which depends only on the galactic astrophysics.
As mentioned in Introduction, if DM decay causes the observations
by ATIC/PPB-BETS, it implies the DM mass $m_{\rm DM}$ should be
about 2 TeV. For $\rho\approx 0.3$ GeVcm$^{-3}$, the decay rate
$\Gamma_{\rm DM}$ should be around $10^{-26}$ sec.$^{-1}$ This
value can be achieved if the DM decays to light leptons through a
dimension 6 operator mediated by a GUT scale massive field ($\sim
10^{16}$ GeV) \cite{strumia}:
\begin{eqnarray} \label{decayrate}
\Gamma_{\rm DM} \sim \frac{m_{\rm DM}^5}{192\pi^3M_{\rm G}^4}\sim
10^{-26}~{\rm sec.}^{-1}
\end{eqnarray}
It might imply that the observations of ATIC/PPB-BETS could be
interpreted as  signals of the GUT scale physics.

If DM with a TeV scale mass was the LSP such as the neutralino or
gravitino, and so supersymmetry (SUSY) breaking soft masses of the
visible sector fields should be heavier than TeV, then the status
of SUSY as a solution of the gauge hierarchy problem in particle
physics becomes weak and a considerable fine-tuning for the Higgs
mass would be unavoidable. Thus, we introduce a Dirac type extra
DM component $\{N,N^c\}$ ($\equiv N_D$) with a TeV scale mass
apart from the (bino-like) LSP, $\chi$. If $N$ and $N^c$ are the
dominant component of DM, the mass of the heavy field mediating DM
decay should be around the GUT scale. However, we have one more DM
component $\chi$, which is the absolutely stable particle.  If the
portion of $\chi$ in the total DM density is large (and so the
portion of $N$ and $N^c$ is small), the mass of the heavy field
mediating DM decay needs to be adjusted such that the flux needed
to explain the positron excess is fixed. That is to say, in
Eq.~(\ref{flux}) the smaller DM number density by a factor
$(M_*/M_G)^4$ ($<1$) can be compensated with a larger decay rate
by assuming the lighter mediator in Eq.~(\ref{decayrate}) such
that the Eq.~(\ref{flux}) remains the same:
\begin{eqnarray}
\frac{\rho}{m_{\rm DM}} \rightarrow \frac{\rho}{m_{\rm
DM}}\times\left(\frac{M_*}{M_G}\right)^4 \quad\quad {\rm for}
\quad\quad M_G\rightarrow M_* \quad {\rm in}\quad \Gamma_{\rm DM}.
\end{eqnarray}
However, we should protect the results of the standard big bang
necleosynthesis.   For the life time of $N$ longer than the age of
the universe $1/\Gamma_{\rm DM}\gtrsim 10^{16}$ sec., the ratio of
the number density of $N$ to $\chi$, $n_N/n_\chi$ is indeed
extremely flexible:
\begin{eqnarray} \label{ratio}
{\cal O}(10^{-10}) ~\lesssim ~ \frac{n_N}{n_\chi}  \quad\quad{\rm
for}\quad 10^{13}~{\rm GeV}\lesssim M_* \lesssim 10^{16}~ {\rm
GeV} .
\end{eqnarray}
Even with an extremely small number density of $N_D$, thus, the
energetic positron excess from the recent experiments can be
easily explained.

In fact, how much $N_D$ was created in the early universe is quite
model dependent.  In Ref. \cite{Hamaguchi}, when DM interacts with
the standard model particles via a GUT suppressed dimension 6
operator, the reheating temperature to produce DM much enough to
explain the energy density $\rho_{\rm DM}\approx 10^{-6}$
GeVcm$^{-3}$ is estimated as $10^{10}$ GeV.  Even if the reheating
temperature is much lower than $10^{10}$ GeV, there exist many
other possibilities to produce $N_D$ sufficiently, depending on
inflationary scenarios.  For instance, $N$ could be non-thermally
created directly from the inflaton decay or indirectly via a
hidden sector field decay. Or $N$ could be coupled to the hidden
sector fields $X$ and $X^c$ with TeV scale masses
%
%
through $W\supset NXX^c$. Then $N$, $N^c$ could be in a thermal
equilibrium state with $X$, $\overline{X}$ (and also $X^c$,
$\overline{X}^c$) by exchanging their scalar partners
$\tilde{X}^c$ ($\tilde{X}$) down to a proper decoupling
temperature, which is defined with hidden sector fields.
%
%
However, we do not specify a possibility in this paper, because we
have two dark matter components and so have extremely large
flexibility for the portions of $n_N/n_\chi$, as mentioned above.

As in the co-annihilation DM scenario of Refs. \cite{HKK,BHKKV},
the {\it electrophilic} coupling of DM,
\begin{equation}
Ne^cE
\end{equation}
is essential also in the DM decay scenario to be consistent with
the PAMELA's observations: They did not observe the anti-proton
excess from the cosmic ray. Here $e^c$ indicates the first family
of the lepton singlet in the MSSM, and $E$ is a newly introduced
heavy lepton singlet with the same electric charge as $e^-$. To
cancel the anomaly, $E$ should be accompanied with $E^c$, whose
electric charge is opposite to that of $E$. They achieve heavy
masses (of order the GUT scale) from the Dirac mass term
$M_EEE^c$.

Of course, one could think the possibility that the singlet $N$
couples also to the quark singlets and extra vector-like heavy
quarks. In this case, however, the coupling with them should be
relatively small such that the hadronic decay rates do not exceed
10 $\%$ for the consistency with the PAMELA's observations
\cite{strumia}.  In this paper we do not introduce such heavy
extra quarks for simplicity of our discussion.

Since the $N$ is required to discriminate leptons and quarks, the
interaction $Ne^cE$ can not be accommodated in the conventional
GUT such as SU(5) and SO(10). It is, however, well embedded in the
flipped SU(5) [$\equiv$ SU(5)$\times$U(1)$_X$] \cite{BHKKV}, which
is a phenomenologically promising GUT. Indeed, flipped SU(5) is
the leptophilic GUT. It means that the leptons, particularly the
lepton singlets are special in flipped SU(5). Since the lepton
singlet $e^c$ remains an SU(5) singlet ($={\bf 1_{5}}$) under
SU(5)$\times$U(1)$_X$, the $N$ ($={\bf 1_0}$) can couple to $e^c$
and $E$ ($={\bf 1_{-5}}$) without being accompanied with quarks,
and $Ne^cE$ ($={\bf 1_01_51_{-5}}$) is invariant under the flipped
SU(5) gauge symmetry.

For dimension 6 dominance in the $N$ decay, the dimension 4 and 5
operators from $Nh_uh_d$, $Nl_ih_u$, $Nl_ih_de^c/M_P$,
$Nl_il_je^c/M_P$, etc. in the superpotential, where $h_u$, $h_d$,
and $l_i$ ($i=1,2,3$) indicate the MSSM Higgs and lepton doublets,
should be removed from the Lagrangian by introducing a proper
symmetry, because they open too fast $N$'s decay channel into
$\chi$ and the standard model leptons ($\Gamma_N\sim
m_N^3/M_{P}^2\sim 10^{-4}$ sec.$^{-1}$, if the dimension 5
operators are dominant.).
The dominant GUT suppressed dimension 6 operator would be
generated from renormalizable operators, in which GUT scale heavy
fields are involved.  The Feynman diagram displaying such induced
dimension 6 operators would satisfy
\begin{equation}
({\rm Number~of~heavy~fermion~propagators}) - ({\rm
Number~of~heavy~mass~insertions}) ~ = ~ 2 .
\end{equation}
We introduce a pair of vector-like heavy SU(2) lepton doublets
$(L,L^c)$ with their Dirac mass term $M_LLL^c$, where $M_L\sim
M_G\sim 10^{16}$ GeV, in the superpotential for the dimension 6
process.

In SUSY theories, a diagram replacing some heavy fermions lines by
the scalar partners' lines is always present. The heavy scalars'
mass squareds are the same as the fermions' upto the SUSY breaking
soft mass squareds of ${\cal O}(m_{3/2}^2)$. Since the
contributions by the scalar propagators are suppressed by $\sim
1/M_G^2$ at low energies, the diagram by the heavy fermions
dominates over the diagram replacing some of them by their scalar
partners.

If the couplings between $E^c$ and the MSSM lepton doublets $l_i$
such as $l_ih_dE^c$ (and also $l_il_jE^c$) were present in the
superpotential, $Nl_ih_de^c/M_E$ (and $Nl_il_je^c/M_E$) could be
induced after integrating out $E$ and $E^c$ from $l_ih_dE^c$
($l_il_jE^c$) and $Ne^cE$. Therefore, $l_ih_dE^c$ and $l_il_jE^c$
also should be disallowed from the superpotential by a symmetry.
Even though they were generated when the symmetry, which forbids
them, is broken, they are still safe only if their coefficients
are suppressed by ${\cal O}(m_{3/2}/M_G)$.

On the other hand, we require the presence of
\begin{equation}
Lh_dE^c \quad {\rm and} \quad m_{3/2}l_1L^c
\end{equation}
in the superpotential in order to make it possible for $N$ (and
also $N^c$) to decay into $\chi e^-e^+$ via the dimension 6
operator. Here $l_1$ stands for the first generation of the lepton
doublet in the MSSM, and $m_{3/2}$ indicates a TeV scale mass
parameter. See the Feynman diagram in Figure 1 for the process
\begin{eqnarray}
N ~(\overline{N^c})\longrightarrow \chi+e^-+e^+  .
\end{eqnarray}
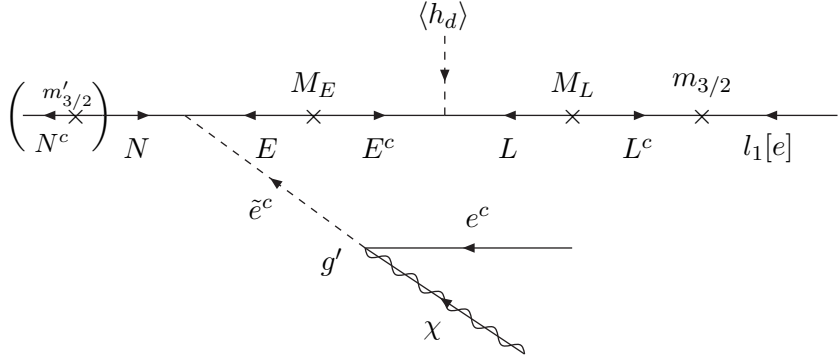
\begin{figure}[t]
\begin{center}
\begin{picture}(332,170)(0,27)

\Text(0,120)[]{$\bigg($} \ArrowLine(23,120)(3,120)
\Text(14,110)[]{\small $N^c$} \Text(23,120)[]{$\times$}
\Text(20,128)[]{\scriptsize $m_{3/2}'$}
\Line(23,120)(33,120)\Text(33,120)[]{$\bigg)$}
\ArrowLine(33,120)(64,120)
\ArrowLine(113,120)(64,120) \ArrowLine(113,120)(162,120)
\ArrowLine(211,120)(162,120) \ArrowLine(211,120)(260,120)
\ArrowLine(310,120)(260,120)

\DashArrowLine(162,150)(162,120){3}
\Text(260,120)[]{\bf $\times$} \Text(260,132)[]{$m_{3/2}$}
\DashArrowLine(132,70)(64,120){3}

\ArrowLine(210,70)(132,70) \ArrowLine(192.06,30)(132,70)
\Photon(192.06,30)(132,70){-2}{6}

\Text(162,158)[]{$\langle h_d\rangle$}
\Text(211,120)[]{\bf $\times$} \Text(113,120)[]{\bf $\times$}
\Text(113,132)[]{$M_E$} \Text(211,132)[]{$M_L$}
\Text(46,108)[]{$N$} \Text(95,108)[]{$E$}
\Text(137.5,108)[]{$E^c$} \Text(186.5,108)[]{$L$}
\Text(235.5,108)[]{$L^c$} \Text(285,108)[]{$l_1[e]$}
\Text(175,82)[]{$e^c$} \Text(93,85)[]{$\tilde{e}^c$}
\Text(158,39)[]{$\chi$} \Text(120,65)[]{$g'$}

\end{picture}
\caption{Dominant diagram of $N$ (and $N^c$) decay: The
dimensionless Yukawa couplings are of order unity. The mass
parameters $M_E$ and $M_L$ are of order the GUT scale.
}\label{fig:Ndecay}
\end{center}
\end{figure}
It is the dominant diagram of $N$ and $N^c$ decay, if $m_N\lesssim
m_{\tilde{e}^c}$.  Its decay rate is estimated as
\begin{eqnarray}
\Gamma_N \approx \frac{m_N^5}{192\pi^3}\times
\left[\frac{g^{'}\langle h_d\rangle m_{3/2}
}{2m_{\tilde{e}^c}^2M_EM_L}\right]^2 \times {\cal O}(y^4) ,
\end{eqnarray}
where $\Gamma_N\sim 10^{-26}~{\rm sec.}^{-1}$ for $m_N\sim 2$ TeV
[$\gtrsim 10\times {\cal O}(m_\chi)$], $M_E\sim M_L\sim 10^{16}$
GeV, and the contributions by the dimensionless Yukawa couplings
${\cal O}(y^4)\sim 1$.

If the selectron $\tilde{e}^c$ is relatively light, $m_{N}\gtrsim
m_{\tilde{e}^c}$, however, the decay channel $N\rightarrow
e^++\tilde{e}^{c*}$ is kinematically allowed, and $\tilde{e}^{c*}$
can be an on-shell particle in Figure 1. Then, the decay rate
becomes enhanced by ${\cal O}(100)$:
\begin{eqnarray}
\Gamma_N\approx\frac{(m_{N}^2-m_{\tilde{e}^c}^2)^2}{16\pi
~m_{N}^3}\left[\frac{g^{'}\langle h_d\rangle m_{3/2}
}{M_EM_L}\right]^2\times {\cal O}(y^4) .
\end{eqnarray}
For $\Gamma_N$ giving $10^{-26}$ sec$^{-1}$, however, $M_E\sim
M_L\sim 10^{16}$ GeV is not much affected.

The $N$ (and also $N^c$) could decay also into the hadrons through
the Higgs of Figure 1. However, such decay modes are much more
suppressed than 10 $\%$. It is because they are 5 body decay
channels with much suppressed phase factors, and the quarks'
(except the top quark) Yukawa couplings at the vertices, from
which quark branches start, are so small.

$L$, $L^c$ and $h_d$ are embedded, respectively, in $\overline{\bf
5}_{-3}$, ${\bf 5}_{3}$ and ${\bf 5}_{-2}$ in flipped SU(5), which
are accompanied with quark singlets. Since $Ne^cE$ ($={\bf
1_01_{5}1_{-5}}$) is still electrophilic and the doublets/triplets
in $\overline{\bf 5}_{2}$, ${\bf 5}_{-2}$ are split by the missing
partner mechanism \cite{Barr82}, however, the decay channels to
hadrons are suppressed also in the flipped SU(5)
model.\footnote{In flipped SU(5), $\bf{10}_1$ and $\overline{\bf
5}_{-3}$ contain $\{d^c,q,\nu^c\}$ and $\{u^c,l\}$, respectively,
where $d^c$ ($u^c$) denotes the quark singlet of $Q_{\rm
e.m.}=1/3$ ($-2/3$), and $q$ ($l$) is the quark (lepton) doublet.
Note that the SU(2) singlets are interchanged by each other,
compared to the Georgi-Glashow's SU(5). Particularly, while the
Majorana neutrino $\nu^c$ is included in the tensor multiplet, the
charged lepton singlet $e^c$ ($={\bf 1_5}$) is the SU(5) singlet.
The MSSM Higgs doublets $h_u$ and $h_d$ are embedded in
$\overline{\bf 5}_{2}$ ($\equiv \overline{\bf 5}_{h}$) and ${\bf
5}_{-2}$ ($\equiv {\bf 5}_{h}$), respectively, in flipped SU(5).
The flipped SU(5) gauge group is broken to the standard model
gauge group by the Higgs fields of the tensor representations
${\bf 10}_H$, ${\bf \overline{10}}_H$. In flipped SU(5), the
doublets/triplets included in $\overline{\bf 5}_{h}$, ${\bf
5}_{h}$ are simply split just through ${\bf 10}_H{\bf 10}_H{\bf
5_{h}}$ and ${\bf \overline{10}}_H{\bf \overline{10}}_H{\bf
\overline{5}_{h}}$.  }

\section{The Model}

The relevant superpotential is composed of $W=W_{N{\rm
decay}}+W_{\rm mass}$, where $W_{N{\rm decay}}$ and $ W_{\rm
mass}$ are, respectively, given by
\begin{eqnarray}
&& W_{N{\rm decay}} = Ne^cE + Lh_dE^c + N^3 + m_{3/2}l_1L^c ,
\label{WNdecay} \\
&& ~~ W_{\rm mass} =  M_{L}LL^c + M_{E}EE^c + m_{N}NN^c ,
\label{Wmass}
\end{eqnarray}
where we drop the dimensionless Yukawa coupling constants for
simplicity. All the dimensionless Yukawa couplings in the above
expressions are tacitly assumed to be {\it of order unity}.
Not introducing $N$ couplings to quarks, we can avoid anti-proton
excess. We assume that the masses of $L$, $L^c$, and $E$, $E^c$,
i.e. $M_L$ and $M_E$ are of order $M_G$ ($\sim 10^{16}$ GeV). On
the other hand, the masses of the DM $N$, $N^c$ should be
constrained to be about 2 TeV in order to explain the observations
of ATIC/PPB-BETS. In fact, the mixing term $l_1L^c$ and the mass
terms of $N$, $N^c$ and also $h_u$, $h_d$, all of which are
proportional to $m_{3/2}$, can not be present in the bare
superpotential due to the U(1)$_R$ symmetry, which will be
presented later. However, via the Giudice-Masiero mechanism
\cite{Giudice88}, the K${\rm \ddot{a}}$hler potential
\begin{equation}
\int d^2\theta d^2\bar{\theta}~
\left[\frac{\Sigma^\dagger}{M_P}\left(\lambda
l_1L^c+\lambda'h_uh_d\right)+{\rm h.c.}\right]
\end{equation}
can induce such supersymmetric mixing term seen in
Eq.~(\ref{WNdecay}), and also the MSSM ``$\mu$ term'', if SUSY is
broken by a nonvanishing VEV of the F component of the singlet
superfield $\Sigma$.  We assume that $\langle F_\Sigma\rangle \sim
m_{3/2}M_P$, where $m_{3/2}\sim$ TeV, $\lambda\sim {\cal O}(1)$,
and $\lambda'\sim {\cal O}(0.1)$. So we have the $\mu$ term of
${\cal O}(100)$ GeV. We will explain later how the Dirac mass term
of $N$ and $N^c$ in Eq.~(\ref{Wmass}) ($m_N\sim 2$ TeV) is
naturally generated. Concerning the gaugino mass terms, the gauge
kinetic functions are involved in their mass generations in
supergravity. We assume that the LSP is the (bino-like) neutralino
with the mass of ${\cal O}(100)$ GeV. It is stable and also a
component of the DM together with $N_D$.

The cubic term of $N$ in Eq.~(\ref{WNdecay}) is introduced such
that the superpartner of $N$, i.e. $\tilde{N}$ promptly decays to
$2\overline{N}$. We just assume that the soft mass of $\tilde{N}$
is heavy enough ($\gtrsim 4$ TeV) for this decay process to open.
Were it not for $N^3$ in the superpotential, $\tilde{N}$ could
remain meta-stable together with $N$. Actually it is not a serious
problem.  But we prefer smaller number of species of DM. By the
SUSY breaking B-term corresponding to the third term in
Eq.~(\ref{Wmass}), $\tilde{N}^c$ can be converted to $\tilde{N}$,
and so $\tilde{N}^c$ also can decay to $2N$ through the $N^3$
term.

\begin{table}[!h]
\begin{center}
\begin{tabular}
%
%
{c|cccccc|ccccc} {\rm Superfields} & $N$ & $N^c$ & $E$ &~ $E^c$ &
$L$ & $L^c$ &~ $\Sigma$ ~
& $e^c$ & $l_1$ & $h_u$ & $h_d$  \\
\hline ${\rm SU(2)}_Y$ & ${\bf 1_0}$ & ${\bf 1_0}$ & ~${\bf
1_{-1}}$ &~ ${\bf 1_1}$ & ${\bf 2_{-1/2}}$ & ${\bf 2_{1/2}}$ &
${\bf 1_0}$ & ${\bf 1_1}$ & ${\bf 2_{-1/2}}$ & ${\bf 2_{1/2}}$ &
 ${\bf 2_{-1/2}}$   \\
$R$ & 2/3 & -4 & ~1/3 & ~5/3
 & 1/3 & 5/3 & 0 & 1 & -5/3 & 0 & 0 \\
$PQ$ & 0 & 2 & 1& -1 & 0 & 0 & 1 & -1 & 1 & 0 & 1  \\
${\cal Z}^{\rm m}_2$ & $+$ & $+$ & $-$ & $-$ & $-$ & $-$ & $+$ &
$-$ & $-$ & $+$ & $+$
%
%
\end{tabular}
\end{center}\caption{The quantum numbers of the superfields. The
superfields written with the capital letters are the newly
introduced fields, which are absent in the MSSM. Except $N$, $N^c$
and the MSSM Higgs fields, the vector-like superfields are all
decoupled from low energy physics due to their heavy masses.
}\label{tab:Qnumb}
\end{table}
%
%
The global symmetry observed in this model is
U(1)$_R\times$U(1)$_{PQ}\times {\cal Z}^{\rm m}_2$, where ${\cal
Z}^{\rm m}_2$ denotes the matter parity (or $R$ parity). The
quantum numbers of the superfields under the symmetry are
displayed in Table~\ref{tab:Qnumb}.
It is straightforward to assign the quantum numbers also to all
other MSSM chiral superfields, which are not presented in
Table~\ref{tab:Qnumb}, so as to admit all the needed $R$ parity
preserving Yukawa terms in the superpotential.

%

In fact, the last terms of Eqs.~(\ref{WNdecay}) and (\ref{Wmass})
violate the U(1)$_R$ and U(1)$_{PQ}$ symmetries. They and soft
SUSY breaking A- and B-terms corresponding to the superpotential
of Eqs.~(\ref{WNdecay}) and (\ref{Wmass}) in the scalar potential
break the $R$ symmetry to $Z_6$. That is to say, SUSY breaking
effects result in U(1)$_R$ breaking into $Z_6$. Since $\Sigma$
carries the unit charge of U(1)$_{PQ}$ symmetry, it is also broken
at the intermediate scale ($\sim \sqrt{m_{3/2}M_P}\sim 10^{10}$
GeV). A proper linear combination of the symmetries could still
remain unbroken, but we will see that it is also broken by VEVs of
other fields.

Among the unwanted terms in the superpotential, which were
discussed in Sec. 2, $Nl_1h_de^c$ and $l_1h_dE^c$ are induced from
the bare K${\rm \ddot{a}}$hler potential: $K\supset \Sigma^\dagger
(Nlh_de^c/M_P^3+lh_dE^c/M_P^2)+{\rm h.c.}$, because they are
consistent with the charge assignments in Table~\ref{tab:Qnumb}.
But their suppression factors $m_{3/2}/M_P^2$ and $m_{3/2}/M_P$
are small enough. However, $Nl_1h_de^c$ and $l_1h_dE^c$
suppressed, respectively, by $m_{3/2}/M_G^2$ and $m_{3/2}/M_G$
rather than $m_{3/2}/M_P^2$ and $m_{3/2}/M_P$ can be generated.
They are exactly what are shown from Figure 1.

Indeed, the diagram in Figure 1 is reminiscent of that explaining
the seesaw mechanism of the neutrino. Integrating out the
superheavy fermions $E$, $E^c$, and $L$, $L^c$ yields the
effective Lagrangian relevant for $N\rightarrow
e^c+\tilde{e}^{c*}$: The equations of motion, $\partial{\cal
L}/\partial E=\partial{\cal L}/\partial E^c=\partial{\cal
L}/\partial L=\partial{\cal L}/\partial L^c=0$, where
$E^{(c)}$,$L^{(c)}$ are the fermionic components of the
corresponding superfields, give $E^c=-\tilde{e}^cN/M_E$,
$E=-\langle h_d\rangle L/M_E$, $L^c=-\langle h_d\rangle E^c/M_L$,
and $L=-m_{3/2}l_1/M_L$, respectively. By inserting them back into
the original Lagrangian, one can get the effective Lagrangian, and
also the effective K${\rm \ddot{a}}$hler potential:
\begin{eqnarray}
{\cal L}_{\rm eff.} = \frac{m_{3/2}}{M_EM_L}h_d\tilde{e}^cl_1N
\quad \subset \quad \int d^2\theta d^2\bar{\theta}~
\left[\frac{\Sigma^\dagger}{M_PM_EM_L}h_de^cl_1N+{\rm h.c.}\right]
.
\end{eqnarray}
Thus, $N$ (and $N^c$) can decay to $e^+e^{c*}$ with the suppressed
amplitude $\sim m_{3/2}\langle h_d\rangle/M_G^2$. By the gauge
interaction ${\cal L}\supset \sqrt{2}g'\tilde{e}^*e^c\chi + {\rm
h.c.}$, $\tilde{e}^{c*}$ eventually decays to $e^-\chi$ as shown
in Figure 1.

\begin{table}[!h]
\begin{center}
\begin{tabular}
%
%
{c|cccc} {\rm Superfields} & $L'$ & $L^{c\prime}$ & $S$ & $\overline{S}$ \\
\hline ${\rm SU(2)}_Y$ &  ${\bf 2_{-1/2}}$ & ${\bf 2_{1/2}}$ &
${\bf 1_0}$
 & ${\bf 1_0}$  \\
$R$  & 1 & 1 & 8/3 & -5/3\\
$PQ$  & 0 & 0 & -1 & 1 \\
${\cal Z}^{\rm m}_2$  & $-$ & $-$ & $+$ & $+$
%
%
\end{tabular}
\end{center}\caption{The quantum numbers of some superfields. They all
are decoupled from low energy physics due to their heavy masses. }
\label{tab:Qnumb2}
\end{table}
%
%
The charge assignments in Table~\ref{tab:Qnumb} forbid the Yukawa
coupling for the electron mass, $l_1h_de^c$ from the bare
superpotential. After U(1)$_R$ and U(1)$_{PQ}$ broken, however,
one can expect the electron mass term is generated.  Let us
consider the following superpotential:
\begin{eqnarray}
W_{\rm elec}=Sl_1L^{c'} + L'h_de^c + M_{L'}L'L^{c\prime}
+\frac{1}{M_P}S^2\overline{S}^2 + \frac{1}{M_P}S^2NN^c,
\label{Welec}
\end{eqnarray}
where the quantum numbers of $S$, $\overline{S}$, and $L'$,
$L^{c'}$ are presented in Table~\ref{tab:Qnumb2}.
The scalar potential derived from the last term of $W_{\rm elec}$,
the A-term corresponding to it, and soft mass terms of $S$,
$\overline{S}$ could allow the minimum, where nonzero VEVs of $S$
and $\overline{S}$ are developed (but $\langle N\rangle=\langle
N^c\rangle=0$), breaking the U(1)$_R$ and U(1)$_{PQ}$ symmetries
completely:
\begin{eqnarray} \label{Svev}
\langle S\rangle\sim \langle \overline{S}\rangle\sim
\sqrt{m_{3/2}M_P}\sim 10^{10}~{\rm GeV} .
\end{eqnarray}
Namely, SUSY breaking triggers the $PQ$ symmetry breaking at the
same energy scale. However the ${\cal Z}^{\rm m}_2$ symmetry is
still unbroken, which plays exactly the role of the matter parity
(or $R$ parity) in the MSSM. The ${\cal Z}^{\rm m}_2$ parity
conservation still prevents the dangerous term $l_il_jE^c$ as well
as $l_il_je^c$ from being induced. Even with the nonzero $\langle
S\rangle$ and $\langle \overline{S}\rangle$, the other unwanted
terms also don't appear at low dimensions.

Once $S$ develops a VEV, the desired mass term of DM $N$ and $N^c$
can be achieved via the last term of Eq.~(\ref{Welec}):
\begin{eqnarray}
m_N = \frac{\langle S^2\rangle}{M_P}\sim 2 ~ {\rm TeV}.
\end{eqnarray}
The electron mass term $(\langle S\rangle/M_{L'})l_1h_de^c$ also
can be generated as seen in Figure 2. It dominates over the
nonrenormalizable term in the bare superpotential, $(\langle
S\rangle/M_P)l_1h_de^c$.
\begin{figure}[t]
\begin{center}
\begin{picture}(270,70)(0,0)

\ArrowLine(1,20)(69,20) \ArrowLine(137,20)(69,20)
\ArrowLine(137,20)(205,20) \ArrowLine(269,20)(205,20)

\DashArrowLine(69,50)(69,20){3} \DashArrowLine(205,50)(205,20){3}

\Text(69,58)[]{$\langle S\rangle$} \Text(205,58)[]{$\langle
h_d\rangle$} \Text(137,20)[]{\bf $\times$}
\Text(137,32)[]{$M_{L^\prime}$} \Text(37,8)[]{$l_1$}
\Text(103,8)[]{$L^{c\prime}$} \Text(171,8)[]{$L^\prime$}
\Text(237,8)[]{$e^c$}

\end{picture}
\caption{Small electron mass generation: The dimensionless Yukawa
couplings are of order unity.}\label{fig:elec.mass}
\end{center}
\end{figure}
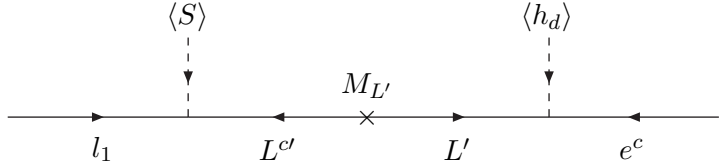
Thus, the correct order of magnitude of the electron mass can be
achieved if $M_{L'}\sim 10^{16}$ GeV:
\begin{eqnarray}
m_e= \frac{\langle S\rangle\langle h_d\rangle}{M_{L'}}\sim
10^{-6}\langle h_d\rangle .
\end{eqnarray}

Expanding $S$ in terms of the axion field $a$ \cite{Kim79,DFSZ81},
$S=(F_aN_{DW}+\rho)e^{ia/(F_aN_{DW})}$, where $F_a$ ($=\langle
S\rangle/N_{DW}\sim 10^{10}~{\rm GeV}/N_{DW}$) is the axion decay
constant and $N_{DW}$ indicates the domain wall number,
$(S/M_{L'})l_1h_de^c$ yields the axion-electron coupling:
\begin{eqnarray}
\frac{m_e}{F_aN_{DW}}~\overline{e}i\gamma_5e ~a ~.
\end{eqnarray}
In fact, if there is a energy loss mechanism by a very weakly
interacting light particle like the axion, it can affect the
resulting luminosity function of the white dwarf. Thus, provided
that the axion is dominantly involved in the cooling mechanism of
white dwarfs, the luminosity function of the white dwarf can be
used to estimate the axion decay constant $F_a$: If the axions
lighter than 1 eV are produced inside white dwarfs, the axion's
coupling to the electron is estimated as $0.7\times 10^{-13}$ for
the best fit of $\chi^2$ \cite{Isern08,BHKKV}, i.e.
\begin{eqnarray} \label{E-Acoupling}
\left|\frac{m_e\Gamma(e)}{F_aN_{DW}}\right| \approx 0.7\times
10^{-13} ,
\end{eqnarray}
where $\Gamma(e)$ denotes the PQ charge of $e$. Thus,
Eq.~(\ref{E-Acoupling}) means $F_aN_{DW}\approx 0.72\times
10^{10}$ GeV for $\Gamma(e)=\pm 1$, which can be coincident with
Eq.~(\ref{Svev}).

\section{Discussion}

The weak scale SUSY is important, because it provides the
resolution of the gauge hierarchy problem and a way to connect the
standard model at low energies and the string theory at high
energies.  If the low energy SUSY should be accepted as a basic
language describing our nature, the DM in the universe should be
understood within this framework.
However, the recently reported observations on excess of energetic
positrons from cosmic ray might be quite embarrassing, because it
is seemingly hard to understand in terms of the conventional MSSM.

In this paper, we present a SUSY model with one more dark matter
component with a TeV scale mass apart from the LSP. The observed
anomaly could be naturally explained by the extra DM's decay
through a dimension 6 operator, which is induced by the
renormalizable operators. The life time of DM ($\sim 10^{26}$
sec.) needed to explain the observed positron flux is caused by
the heavy masses ($\sim 10^{16}$ GeV) of the mediators involved in
the decay process. Since this model permits the possibility that
the field spectrum below the GUT scale can coincide with that of
the MSSM except for $N_D$, the gauge coupling unification in the
MSSM could be still maintained. Therefore, we could rescue the
string models realizing the MSSM with sin$^2\theta_W=3/8$
\cite{stringMSSM}, because a lot of neutral singlets are easily
found in string models.

\acknowledgments{ \noindent The author is supported by the FPRD of
the BK21 program, in part by the Korea Research Foundation, Grant
No. KRF-2005-084-C00001 and the KICOS Grant No.
K20732000011-07A0700-01110 of the Ministry of Education and
Science of Republic of Korea. }

\end{document}